\documentclass[aps,prb,twocolumn,showpacs,superscriptaddress,endfloats]{revtex4}
\usepackage{graphicx}

\begin{document}

\title{Canted ferromagnetism in RuSr$_2$GdCu$_2$O$_8$}
\author{Kohji~Nakamura}
\email[Email address: ]{kohji@phen.mie-u.ac.jp}
\affiliation{Department of Physics Engineering, Mie University, Tsu, Mie 514-8507, Japan}
\affiliation{Department of Physics and Astronomy, Northwestern University, Evanston,
IL60208}
\author{A.~J.~Freeman}
\affiliation{Department of Physics and Astronomy, Northwestern University, Evanston,
IL60208}
\date{\today}

\begin{abstract}
First principles calculations using the full-potential linearized augmented
plane wave (FLAPW) method including intra-atomic noncollinear magnetism have
been performed to determine the magnetic structures of RuSr$_{2}$GdCu$_{2}$O$%
_{8}$. The magnetism clearly arises from the RuO$_{6}$ octahedra where the
moments on neighboring Ru sites order antiferromagnetically but cant 
perpendicular to the AFM axis - and so induce a weak
ferromagnetism. 
The projected Ru moments along the AFM and FM axes result in magnetic
moments of 1.16 and 0.99$\mu _{B}$ respectively. 
The results are consistent with the possible coexistence of canted
ferromagnetism and superconductivity in the RuSr$_{2}$GdCu$_{2}$O$_{8}$ -
inferred from experiments.
\end{abstract}

\maketitle


The coexistence of magnetism and superconductivity on a microscopic scale
has recently been reported in RuSr$_{2}$RCu$_{2}$O$_{8}$ (R=Gd, Eu and Y).%
\cite{99Ber,00Wil,01Tok} These materials have an analogy to the
superconductivity in the two-dimensional high $T_{c}$ cuprate
superconductors associated with the Cu~$e_{g}$ states of the CuO$_{2}$
layers. The superconductivity appears below a superconducting transition
temperature, $T_{c}=$30-50~K. The magnetism arises from the Ru~$t_{2g}$
states of the RuO$_{6}$ octahedra, which do not produce any significant
effects on the superconductivity since the Ru~$t_{2g}$ electrons do not
couple to the Cu~$e_{g}$ states regardless of the ordering of the magnetic
Ru moments.\cite{99Pic,01Nak}

Although the Ru moments order magnetically below a magnetic transition
temperature, $T_m$=130-150~K, the type of ordering still remains
controversial. An earlier report of a homogeneous ferromagnetic (FM)
ordering of the Ru moments by dc magnetization and muon spin rotation
experiments\cite{99Ber} has been brought into a question by neutron
diffraction experiments,\cite{00Lyn} suggesting an antiferromagnetic (AFM)
G-type ordering where nearest neighbors of the Ru moments along all three
crystallographic axes are coupled antiferromagnetically. First-principles
calculations also demonstrated that the AFM ordering is energetically
favored over the FM one within the collinear magnetic structures.\cite{01Nak}
Nevertheless, recent magnetization and magnetic resonance experiments\cite%
{01But1,01But2,00Fai} have provided clear evidence of a weak ferromagnetism.
To account for complexity in the magnetism, noncollinear magnetism such as
canting of the Ru moments has been proposed,\cite{00Wil,01Jor} to induce a
ferromagnetic component.

Here, we determine from first-principles the magnetic structures in RuSr$_{2}
$GdCu$_{2}$O$_{8}$ by using the highly precise full-potential linearized
augmented plane-wave (FLAPW) method\cite{81Wim} that includes intra-atomic
noncollinear magnetism which can describe the canting of the Ru moments.
Indeed, we find that the moments on neighboring Ru sites order
antiferromagnetically but cant 
perpendicular to the AFM axis - and so induce a weak
ferromagnetism. 

In the calculations, we employed a crystal structure with the $P4/mbm$ space
group determined by neutron diffraction experiments.\cite{00Chm} This
structure is similar to that of YBa$_2$Cu$_3$O$_7$, where Y, Ba and Cu
(chain atoms) are replaced by Gd, Sr and Ru, respectively. Ru lies at a
six-fold coordinated position in the octahedron composed of six neighboring
oxygens (four O$_{\mathrm{Ru}}$ and two O$_{\mathrm{apical}}$) while Cu lies
in a five-coordinated position (four O$_{\mathrm{Cu}}$ and one O$_{\mathrm{%
apical}}$). The RuO$_6$ octahedra are rotated by about 14$^\circ$ around the
c-axis, which leads to a significant modification in the electronic and
magnetic structures; hence the magnetism of the Ru is sensitive to the
structural distortion.\cite{01Nak}

The FLAPW calculations were performed based on the local spin density
approximation, L(S)DA, with the Hedin-Lundquist exchange-correlation.\cite%
{71Hed,72Bar} Although the effects of electronic correlation in a strongly
correlated system may be taken into account within a scheme such as LDA+U,
RuSr$_{2}$GdCu$_{2}$O$_{8}$ shows metallic character even in the RuO$_{2}$
layers, which causes that effect to be weak and so may not alter our
results. The intra-atomic noncollinear magnetism formalism\cite{72Bar,88Kub}
was incorporated into the FLAPW method with no shape approximation for the
magnetization density\cite{96Nod,02Nak}; in this, the density functional
theory is treated with a density matrix with 2$\times $2 components of the
charge and magnetization density. Our approach allows the magnetic moment
direction as well as the magnitude to vary continuously all over space,
i.e., no shape approximation for the magnetization. The plane-wave was
augmented with a spin-independent LAPW basis at the muffin-tin boundary.
Although this approximation loses freedom compared with having a separate
spin-up and spin-down LAPW basis, as is done in calculations for collinear
magnetism, we confirmed that the accuracy was not degraded.\cite{chkncm} The
calculations were carried out without spin-orbit coupling (SOC),
since the SOC induced energies,\cite{99Wu} such as magneto-crystalline
anisotropy energy that determines the easy direction, is expected to be much
smaller than those of interest here.

The self-consistent calculations were started with an initial magnetization
density depicted in Fig.~\ref{fig1}(b), which is a magnetic structure
similar to a C-type AFM structure (Fig.~\ref{fig1}(a)) but with the Ru
moments slightly canted 
perpendicular to the AFM axis, i.e., to the FM axis direction in
the figure. Here, the moment directions throughout the present paper are
defined in a spin space since no SOC is taken into account. Although the
neutron diffraction experiments\cite{00Lyn} revealed an AFM G-type ordering,
we employed the C-type one in order to reduce the large computational effort
this would entail. (The G-type ordering requires doubling the unit cell of
the C-type ordering.) This is justified because the moment alignment along
the c-axis is less important than that along the a-axis since the distance
between the neighboring Ru atoms along the c-axis (11.56~\AA ) is
significantly greater than that along the a-axis (3.84~\AA ).

\begin{figure}
\caption{Schematic magnetic ordering of Ru and Gd moments in (a) collinear
AFM and (b) noncollinear AFM structures of RuSr$_2$GdCu$_2$O$_8$, where the
Cu, O$_{\mathrm{Ru}}$, O$_{\mathrm{Cu}}$ and O$_{\mathrm{apical}}$ atoms are
not given. The ordering of the Ru moments in (b) is similar to a C-type AFM
ordering in (a) but the moments cant slightly out of their original
direction, i.e., along the FM axis direction. Note that the moment
directions are defined in a spin space since SOC is not taken into account.}
\label{fig1}
\end{figure}

The calculated spin magnetization density in the RuO$_{2}$ layer for the
(110) and (001) planes is shown in Fig.~\ref{fig2} (a) and (b),
respectively. The magnetism is clearly visible in the Ru, O$_{\mathrm{Ru}}$
and O$_{\mathrm{apical}}$ ions. The moments on the neighboring Ru sites
order antiferromagnetically but their moments cant along the FM axis
direction, which induces a ferromagnetic component of the magnetic moments.
The canting of the Ru moments appears in the $t_{2g}$ orbitals. The
magnetization in O$_{\mathrm{apical}}$ is found to correlate with that in
the Ru, where the O$_{\mathrm{apical}}$~$p_{x(y)}$ moments tilt in the same
way as those in the Ru~$d_{xz(yz)}$ orbitals. The O$_{\mathrm{Ru}}$ moments
are also induced but point only along the FM axis direction, which
correlates with the Ru~$d_{xy}$ states, as seen in Fig.~\ref{fig2}(b). The
Ru, O$_{\mathrm{Ru}}$ and O$_{\mathrm{apical}}$ magnetic moments inside the
muffin-tin spheres are given in Table~\ref{table1}, where the moments are
projected along the AFM and FM axes. It is striking that the projected Ru
moments along the AFM and FM axes result in magnetic moments of 1.16 and
0.99~$\mu _{B}$, respectively, which have similar values to that observed by
experiments, 1.18~$\mu _{B}$ when measured as the AFM structure by neutron
diffraction\cite{00Lyn} and 1~$\mu _{B}$ as the FM value by 
magnetization.\cite{99Ber} 
However, a direct comparison of the
projected FM moment with the experimetal one would be difficult, since
complicated behaviors, such as a spin-flop transition, accompany the high
field experiments. Neutron diffraction\cite{00Lyn} demonstrated that the
fields exceeding 0.4~T gradually enhance the FM intensity but decrease the
AFM intensity; the highest field of 7~T results in a FM moment of 1.4~$\mu
_{B}$ with no significant AFM intensity. In such a high field, the Ru
moments tend to align in a collinear FM state, leading to the enhancement of
the FM moments, which roughly agrees with the calculated moments ($\sim $1.5~%
$\mu _{B}$) in the collinear FM state.\cite{01Nak} Clearly, further
investigations including the field dependence are necessary to fully
describe the different experimental observations. 

\begin{figure}
\caption{Spin magnetization density in the RuO$_2$ layer for the (110) and
(001) planes of RuSr$_2$GdCu$_2$O$_8$, where the moment direction and
magnitude are represented by arrow and the size, respectively.}
\label{fig2}
\end{figure}

\begin{table}
\caption{Calculated magnetic moments, $m$ (in $\protect\mu_B$), in the MT
spheres of Ru, O$_{\mathrm{Ru}}$ and O$_{\mathrm{apical}}$ of RuSr$_2$GdCu$_2
$O$_8$; here $m_{\mathrm{AFM}}$ and $m_{\mathrm{FM}}$ are projected moments
along the AFM and FM axes in Fig.~1. Ru and O$_{\mathrm{apical}}$ have two
sublattice sites. }
\label{table1}%
\begin{ruledtabular}
\begin{tabular}{lccc}
& $m_{\mathrm{AFM}}$ & $m_{\mathrm{FM}}$ & $|m|$ \\ \hline
Ru & $\pm$1.16 & 0.99 & 1.53 \\ 
O$_{\mathrm{Ru}}$ & 0.00 & 0.08 & 0.08 \\ 
O$_{\mathrm{apical}}$ & $\pm$0.08 & 0.07 & 0.11\\
\end{tabular}
\end{ruledtabular}
\end{table}

By introducing intra-atomic noncollinear magnetism, we found that the
calculated total energy is only 10~meV/cell 
lower than that of the collinear AFM state. Thus, the canting of the Ru
moments is energetically favored in the system. However, since the total
energy difference is so small compared with $T_{m}$, there would only be a
short-range ferromagnetic ordering. This may be a reason why experiments,
such as the neutron diffraction\cite{00Lyn,00Chm} in the low-fields
up to 0.4~T, could not clearly detect the ferromagnetic moments within
their experimental sensitivity.

In order to discuss the Ru magnetism, we first consider the collinear AFM
case and present the density of states (DOS) of the Ru~$t_{2g}$ in Fig.~\ref%
{fig3}(a), where the $x$ and $z$-axes are chosen as directions to the
neighboring O$_{\mathrm{Ru}}$ and O$_{\mathrm{apical}}$ sites, respectively.
The gray regions indicate the weight of the majority spin states. The
magnetism of the Ru is dominated by antibonding $t_{2g}$ states.\cite{01Nak}
The majority spin $d_{xy}$ and $d_{xz(yz)}$ states on the Ru site can
hybridize with the minority spin states on the neighboring Ru sites through O%
$_{\mathrm{Ru}}$ $p_{x(y)}$ and $p_z$ orbitals by a superexchange mechanism.
The majority spin $d_{xy}$ and $d_{xz(yz)}$ states are almost fully
occupied; therefore, the charge configuration of the Ru is close to $t_{2g}^3
$ (Ru$^{5+}$) with a high spin state, which prefers an antiferromagnetic
alignment of their moments.\cite{59Kan} (Note that itinerant electrons are
partially occupied in the minority spin Ru~$d_{xy}$ states, which creates an
electron pocket at the $\mathit{\Gamma}$ point in the Fermi surface, not
shown, and leads to metallic RuO$_2$ layers.)

\begin{figure}
\caption{Density of states (DOS) of Ru $t_{2g}$ states for (a) collinear AFM
and (b) noncollinear AFM structures. The gray regions indicate the weight of
their majority spin states.}
\label{fig3}
\end{figure}

When the intra-atomic noncollinear magnetism is introduced, however, the
admixture of the spin up and down states leads to an additional
hybridization between the neighboring Ru ions, and results in the wider
bandwidth of the $t_{2g}$ states seen in Fig.~\ref{fig3}(b). Although there
is an admixture of the spin up and down states by introducing intra-atomic
noncollinear magnetism, the spin-projected DOS along the average moment
direction is plotted in the figure. The majority spin Ru~$d_{xz(yz)}$ bands
become more dispersive and cross the Fermi level ($E_{F}$) while the
minority spin Ru~$d_{xy}$ bands are more occupied by itinerant electrons;
this causes the system to be more metallic in character and close to a Ru$%
^{4+}$ state with a low spin state configuration - as expected from
experiments.\cite{01But1,01Liu} Hence, a double exchange interaction due to
the itinerant electrons that induces a weak ferromagnetism is promoted, and
the magnitude of the Ru moments is significantly reduced from that expected
from Hund's rule. Of course, the specification of the Ru valence states is,
however, not exact due to the metallic character. It should be noted that
this noncollinear magnetism arises from the band effects just discussed but
does not have its origin in the SOC, such as via Dzyaloshinsky-Moriya
interactions.\cite{60Mor}

Finally, we comment the electronic structure of the CuO$_{2}$ bilayer.
Even if the Ru moments cant, the Cu~$e_{g}$ band structure
that is responsible to the high $T_C$ superconductivity
was found to be almost the same as those predicted 
by previous calculations of the collinear AFM state.\cite{01Nak} 
This is because of an unique electronic structure of the
layered Ru~$t_{2g}$ and Cu~$e_{g}$ states separated by O$_{\mathrm{apical}}$~%
$p$ orbitals. The Ru~$t_{2g}$ states couple to the O$_{\mathrm{apical}}$~$%
p_{x(y)}$ orbitals which do not couple to the Cu~$e_{g}$ states.\cite%
{99Pic,01Nak} 
Therefore, as previously demonstrated,\cite%
{99Pic,01Nak} the strong hybridized Cu-O $dp\sigma $ orbitals, which show
nesting Fermi surface features similar to those in the high $T_{c}$ cuprate
superconductors, will give rise to anomalous behavior of the electronic
properties arising from singularities in the generalized susceptibility.
This is consistent with the possible coexistence of canted ferromagnetism
and superconductivity in RuSr$_{2}$GdCu$_{2}$O$_{8}$. 

In summary, first principles FLAPW calculations including intra-atomic
noncollinear magnetism were performed to determine the magnetic structure of
RuSr$_2$GdCu$_2$O$_8$. The magnetic moments on the neighboring Ru sites
order antiferromagnetically but cant 
perpendicular to the AFM axis. From the canting of the Ru moments,
a double exchange interaction is exerted via itinerant $t_{2g}$ electrons
which can travel through the neighboring O~$p$ states. 
The results also suggest the possible coexistence of canted
ferromagnetism and superconductivity in RuSr$_2$GdCu$_2$O$_8$.

Work at Northwestern University supported by the U.~S. Department of Energy,
Office of Science under Grant No.DE-FG02-88ER45372.


\begin{thebibliography}{99}
\bibitem{chkncm} To confirm the accuracy, we performed FLAPW calculations
using the spin-dependent and spin-independent LAPW basis for the collinear
AFM C-type structure of RuSr$_{2}$GdCu$_{2}$O$_{8}$\ (cf.,Fig. 1 (a)) and
found that both results are consistent. The calculated Ru and Gd magnetic
moments showed the same values 1.57(-1.53) and 6.9~$\mu _{B}$, respectively,
in both cases.
\end{thebibliography}
\end{document}